\documentclass[prb,showpacs,twocolumn,amsmath,amssymb,groupedaddress]{revtex4}
\usepackage{graphicx}
\usepackage{enumitem}
\usepackage{bbold}
\usepackage[colorlinks,bookmarks=false,citecolor=red,linkcolor=blue,urlcolor=blue]{hyperref}

\allowdisplaybreaks

\begin{document}

\title{Dirac cones and mass terms in bosonic spectra}
\author{P. Sathish Kumar}
\affiliation{The Institute of Mathematical Sciences, HBNI, C I T Campus, Chennai 600 113, India}
\author{R. Ganesh}
\email{ganesh@imsc.res.in}
\affiliation{The Institute of Mathematical Sciences, HBNI, C I T Campus, Chennai 600 113, India}

\date{\today}

\begin{abstract}
The notion of Dirac cones, wherein two or more bands become degenerate at a certain momentum, is the starting point for the study of topological phases. Dirac cones have been thoroughly explored in fermionic systems such as graphene, Weyl semimetals, etc.  
The underlying mathematical structure in these systems is a Clifford algebra -- a rule for identifying sets of matrices that span the Hamiltonian. This structure allows for the identification of suitable `mass' terms to open band gaps. In this article, we extend these ideas to bosonic systems. Due to the pseudo-orthogonal nature of eigenvectors, the algebra of matrices takes a very different form. Taking the honeycomb XY ferromagnet as a prototype, we show that a Dirac cone emerges in the magnon spectrum. A gap can be opened by a suitable mass term involving next-nearest neighbour interactions. We next construct a one-dimensional ladder model with triplon excitations. Using the new Clifford algebra, we define winding number as a topological invariant. In analogy with the Su-Schrieffer-Heeger model, topological transitions occur when the band gap closes, leading to the appearance (or disappearance) of protected edge states. Our results suggest a new route to studying band touching and band topology in bosonic systems.
\end{abstract}
                                 
\maketitle

\section{Introduction}
The rise of topological insulators stems from discoveries in electronic band structures. In contrast, there has been a recent surge of interest in topological phases of bosonic systems\cite{Onose2010,Matsumoto2011,Hoogdalem2013,Zhang2013,Romhanyi2015}.
 As bosonic particles are typically charge-neutral and weakly interacting, they hold promise for edge state transport with long coherence times. However, due to their bosonic character, the nature of the eigenvectors is fundamentally different. While fermionic band structures are well understood, we do not yet have a clear understanding of bosonic systems and their topological principles. In this article, we show that two central aspects of fermionic band topology -- Dirac cones and the notion of a Clifford algebra -- can be extended to bosonic systems. 

The field of topological insulators arose from the study of Dirac cones and mass terms. The seminal discoveries of Haldane\cite{Haldane1988} and Kane and Mele\cite{KaneMele2005} were made in the context of electrons living on a honeycomb lattice. This system provides a two dimensional analogue of the Dirac equation. In particular, it allows for Dirac cones -- wherein two bands touch at a single point in momentum space. A band gap can be opened by introducing a suitable `mass' term. The Dirac cone Hamiltonian and the mass term constitute a `Clifford algebra', a mathematical rule for identifying sets of matrices. This structure also underlies more advanced discoveries such as quadratic band touching points\cite{Janssen2015}, deconfined criticality\cite{Moon2016}, etc. 

As we show below, the Clifford algebra structure does not carry over to bosonic systems. At the same time, we have an ever growing number of examples of Dirac-like band touching points in bosonic spectra. Examples include Dirac cones in phonons\cite{Huang2012,Yu2015,Yu2016,Gao2016}, photons\cite{Sakoda2012,Chan2012,He2015,Yi2016}, magnons\cite{Owerre2016,Fransson2016,Kim2016,Owerre_SciRep_2017} and triplons\cite{Romhanyi2015,McClarty2017,Joshi2017}. More recently, there has been an explosion of interest in Weyl points in magnonic band structures\cite{Li2016,Mook2016,Kangkang2017,Li2017,Ying2017,Jian2017}. By analogy with the fermionic case, such systems with band touching points should be excellent starting points for topological physics. In Sec.~\ref{sec.fermi} below, we first review the physics of band touching points in fermionic band structures. In Sec.~\ref{sec.bose}, we review bosonic Dirac points taking the example of magnons in the honeycomb XY ferromagnet. We describe a new Clifford algebra-like structure that arises. We identify a suitable mass term and discuss consequences for topology. 
In Sec.~\ref{sec.triplon}, we consider a ladder system with triplon excitations constituting a one-dimensional realization of this algebra. This allows us to define `winding number' as a topological invariant with non-trivial systems developing edge states. We conclude with a discussion about applications to other magnetic systems. 

\section{Clifford algebras and topology in fermionic systems}
\label{sec.fermi}
We review the key features of Dirac cones in fermionic systems here. We will build analogous structures for bosons in the following sections. 
It can be said that the physics of topological insulators arose from the study of fermions hopping on a honeycomb lattice, a model with immediate relevance to graphene. 
This model gives rise to two bands which touch at two points in the Brillouin zone. At half-filling, this leads to point-like Fermi surfaces with two `Dirac points'. The Hamiltonian describing the (spinless) fermions takes the form
\begin{eqnarray}
H = \sum_\mathbf{k} \Psi_{\mathbf{k}}^\dagger H_\mathbf{k}  \Psi_{\mathbf{k}},  
\label{eq.Hamiltonian}
\end{eqnarray}
where $\Psi_{\mathbf{k}}=\left( \begin{array}{cc}c_{\mathbf{k},A} & c_{\mathbf{k},B} \end{array}\right)^T$ is the vector of annihilation operators, with $A/B$ denoting the two triangular sublattices that constitute the honeycomb lattice. The operators here satisfy fermionic anticommutation relations, i.e., $\{ (\Psi_{\mathbf{k}})_i,(\Psi_{\mathbf{k'}}^\dagger)_j \} = \delta_{\mathbf{k},\mathbf{k}'}\delta_{i,j}$. 
Up to an overall shift, the Hamiltonian is a $2\times 2$ matrix with a simple form.
\begin{eqnarray}
H_\mathbf{k} = f_x(\mathbf{k}) \sigma_x + f_y(\mathbf{k}) \sigma_y +m(\mathbf{k}) \sigma_z,
\label{eq.HPauli}
\end{eqnarray}
where $\sigma_{x/y/z}$ are Pauli matrices. With only nearest neighbour hopping, the coefficient of $\sigma_z$ vanishes uniformly with $m(\mathbf{k})=0$. In the vicinity of the Brillouin zone corner, the other coefficients take a simple form with $f_x\sim k_x$ and $f_y\sim k_y$, where $k_x$ and $k_y$ denote displacements from the $K$ point. With a suitable perturbation, a non-zero $m(\mathbf{k}) $ may be introduced.

This Hamiltonian exemplifies the notion of a Clifford algebra, a set of matrices satisfying the following two conditions:
(i) the matrices must each square to the identity matrix and (ii) they must anticommute with one another. Pauli matrices constitute the simplest example, forming a three-element Clifford algebra. 
Using these properties, we can immediately identify the eigenvalues of the Hamiltonian by the following argument.

To diagonalize the Hamiltonian, we need a suitable unitary transformation, i.e., $U_\mathbf{k}^\dagger H_\mathbf{k} U_\mathbf{k} = \mathrm{Diag}\{E_1(\mathbf{k}),E_2(\mathbf{k})\}$, where $U_\mathbf{k}^\dagger U_\mathbf{k} = \sigma_0$. The transformation matrix should be unitary so as to preserve fermionic anticommutation relations.
The eigenvalues can be directly deduced without finding the explicit form of $U_\mathbf{k}$, by considering the square of the Hamiltonian. Using (i) and (ii) above, we find 
\begin{eqnarray}
H_{\mathbf{k}}^2 \sim \left\{ f_x^2(\mathbf{k}) + f_y^2 (\mathbf{k})+ m^2(\mathbf{k}) \right\} \sigma_0.
\label{eq.Hsigma}
\end{eqnarray}
We note that the unitary matrix that diagonalizes $H_\mathbf{k}$ will also diagonalize $H_\mathbf{k}^2$, since 
\begin{equation}
U_\mathbf{k}^\dagger H_\mathbf{k}^2 U_\mathbf{k} = U_\mathbf{k}^\dagger H_\mathbf{k}U_\mathbf{k} U_\mathbf{k}^\dagger H_\mathbf{k} U_\mathbf{k} = \mathrm{Diag}\{ E_1^2(\mathbf{k}),E_2^2(\mathbf{k})\}.
\label{eq.Hsquare}
\end{equation}
Here, we have inserted an identity matrix in the form of $U_\mathbf{k} U_\mathbf{k}^\dagger$. Thus, the eigenvalues of $H_\mathbf{k}^2$ are trivially related to those of $H_\mathbf{k}$. While $H_\mathbf{k}$ may be difficult to diagonalize, the eigenvalues of $H_{\mathbf{k}}^2$ are immediately found from Eq.~\ref{eq.Hsigma}. We deduce that the eigenvalues of $H_\mathbf{k}$ are $E_{1/2}(\mathbf{k})=\pm \sqrt{f_x^2 (\mathbf{k})+ f_y^2 (\mathbf{k})+ m^2(\mathbf{k})}$, without having to find $U_\mathbf{k}$. In the vicinity of the Dirac point, we have $E_{1/2}\approx \pm \sqrt{k_x^2 + k_y^2 + m_D^2} $ where $m_D$ is the value of $m(\mathbf{k})$ at the Dirac point. We see that we have a band gap of $2 m_D$; we identify the $\sigma_z$ term in the Hamiltonian as a `mass' term that opens a band gap.   

In the honeycomb lattice system, there are two distinct Dirac points. To open a band gap, mass terms must be introduced at both. This can be done in two well-known ways (without extending the unit cell): (i) The Semenoff mass arises from a sublattice potential which amounts to two mass terms with the same sign, i.e., $\mathrm{sign}(m_1)=\mathrm{sign}(m_2)$\cite{Semenoff1984,Hunt2013}. (ii) In contrast, the Haldane mass arises from a complex next-nearest neighbour hopping, giving rise to mass terms of opposite sign, i.e., $\mathrm{sign}(m_1)\neq\mathrm{sign}(m_2)$\cite{Haldane1988}.

\section{Dirac cones in a bosonic Hamiltonian}
\label{sec.bose}
\begin{figure}
\includegraphics[width=3in]{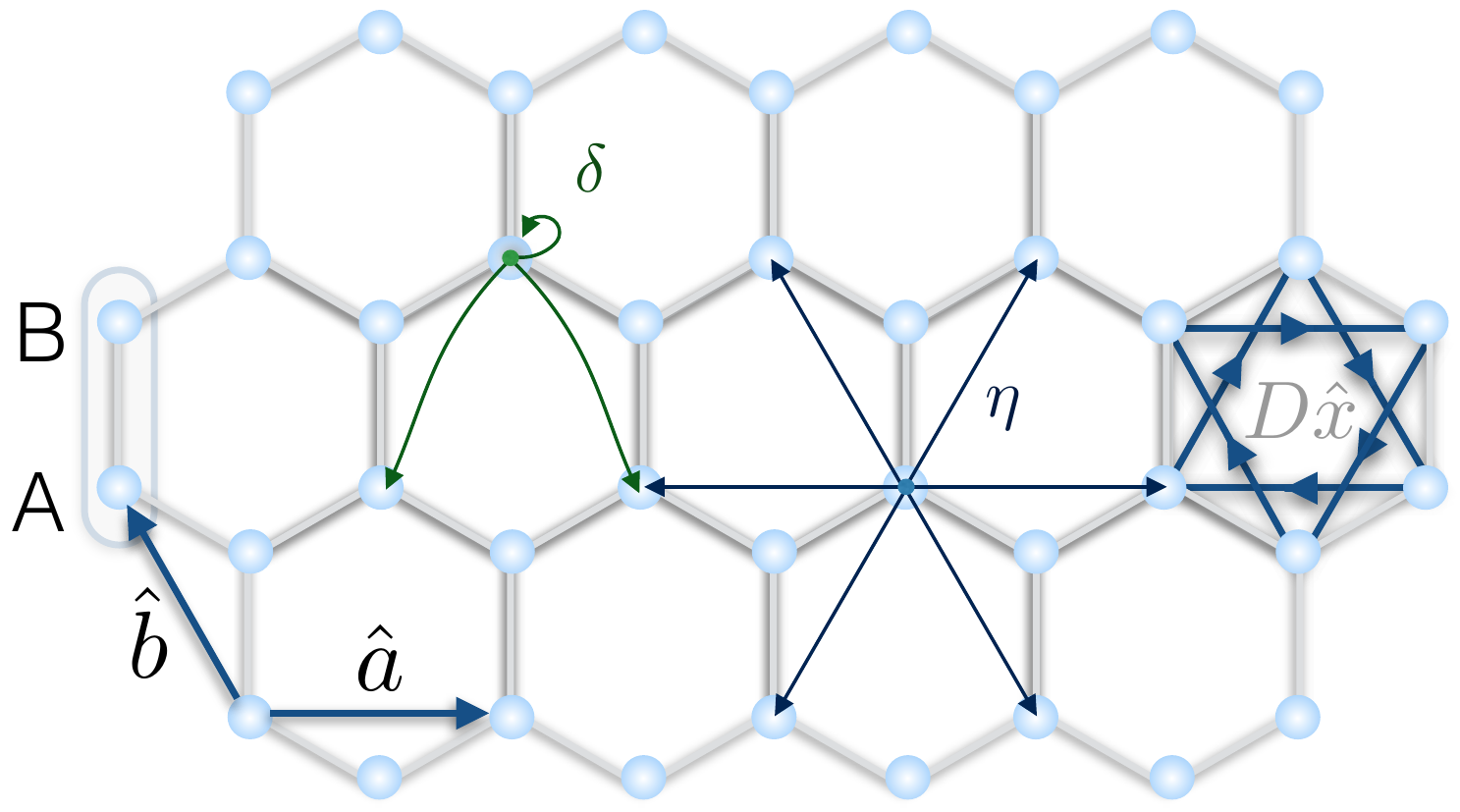}
\caption{The honeycomb lattice with the two-site unit cell. The primitive lattice vectors, $\hat{a}$ and $\hat{b}$, span each triangular sublattice. The three nearest neighbour vectors, $\delta$, are shown as green arrows. The six next-nearest neighbour vectors, $\eta$, are shown as blue arrows. DM couplings that open a gap (see text) are shown in one hexagon.}
\label{fig.lattice}
\end{figure}

The simplest non-trivial example of a Dirac cone in a bosonic system occurs in the honeycomb lattice XY ferromagnet. We consider the Hamiltonian
\begin{eqnarray}
\nonumber H_{XY} &=& -J\sum_{i} \sum_{\delta} \left[ S_{i,A}^x S_{i+\delta,B}^x + S_{i,A}^y S_{i+\delta,B}^y \right]\\
 &-& h \sum_{i} \left[S_{i,A}^x  + S_{i,B}^x \right],
\end{eqnarray}
where the index $i$ runs over all unit cells of the honeycomb lattice, shown in Fig.~\ref{fig.lattice}. The three nearest neighbours of a given A-sublattice site are denoted by $(i+\delta,B)$, the B site of the unit cell at $(i+\delta)$, with $\delta$ taking three possible values. The magnetic field $h$ breaks in-plane rotation symmetry and selects a ground state with ferromagnetic moment along the X direction. The excitations about this state are spin waves or magnons, with the Hamiltonian,
\begin{eqnarray}
H =   JS\sum_{\mathbf{k}}{}^{'} \Phi_\mathbf{k}^\dagger H_{\mathbf{k}}^b \Phi_\mathbf{k} + \mathrm{const.}
\label{eq.Hamiltonianbosons}
\end{eqnarray} 
The primed summation signifies that if $\mathbf{k}$ is included in the sum, $-\mathbf{k}$ must be excluded. The vector of operators $\Phi_\mathbf{k}$ and the Hamitonian matrix are given by
\begin{eqnarray}
\nonumber
 \Phi_\mathbf{k} &=& \left(  \begin{array}{cccc}
a_{\mathbf{k}} &
b_{\mathbf{k}} &
a_{-\mathbf{k}}^\dagger &
b_{-\mathbf{k}}^\dagger
\end{array}
 \right)^T, \\ 
 H_{\mathbf{k}}^b &=&
\left( 
\begin{array}{cccc}
 3+h/J&-\epsilon(\mathbf{k}) &0 & \epsilon(\mathbf{k})\\ 
  -{\epsilon}^{*}(\mathbf{k})&3+h/J &{\epsilon}^{*}(\mathbf{k}) &0 \\
  0 & \epsilon(\mathbf{k}) &3+h/J & -\epsilon(\mathbf{k})\\  
  {\epsilon}^{*}(\mathbf{k})&0 &-{\epsilon}^{*}(\mathbf{k})   & 3+h/J
 \end{array}\right)\!\! ,
 \end{eqnarray} 
where the bosonic operators are defined as $a_\mathbf{k} = \sum_{i\in A} a_i e^{i\mathbf{k}\cdot \mathbf{r}_i}$ and similarly for $b_\mathbf{k}$. The operators $a_i^\dagger$ and $b_i^\dagger$ create a spin excitation on the A and B sites of the unit cell labelled by $i$.
 We have defined $\epsilon_\mathbf{k}=\frac{1}{2}\sum_\delta e^{i\mathbf{k}\cdot \mathbf{\delta}}$.

There are several important differences vis-\`a-vis fermionic systems such as that described by Eqs.~\ref{eq.Hamiltonian},\ref{eq.HPauli}. The key differences are:
\begin{enumerate}[label=(\alph*)]
\item The Hamiltonian need only be defined over half the Brillouin zone due to the primed summation over $\mathbf{k}$.
\item The Hamiltonian contains pairing terms, e.g., $a_{\mathbf{k}}^\dagger b_{-\mathbf{k}}^\dagger$. This may also apply to fermionic systems in the presence of superconductivity.
 \item The elements of $\Phi_\mathbf{k}$ satisfy bosonic commutation relations, i.e., $[(\Phi_\mathbf{k})_i, (\Phi_{\mathbf{k}'}^\dagger)_j] = \delta_{\mathbf{k},\mathbf{k}'} (\mu_C)_{i,j}$, where $\mu_C = \mathrm{Diag}\{1,1,-1,-1\}$ is the commutation matrix.
\end{enumerate}

The most important difference in bosonic systems is (c) above. On account of the bosonic commutation relations, the matrix that diagonalizes the Hamiltonian can no longer be unitary (as it will not preserve the commutation relations). We require a `pseudo-unitary' matrix, $W_{\mathbf{k}}$, which satisfies the following properties\cite{Blaizot1986}:
\begin{eqnarray}
W_{\mathbf{k}}^\dagger  H_{\mathbf{k}}^b W_\mathbf{k} &=& \mathrm{Diag}\{E_1,E_2,E_3,E_4\}, \\
W_{\mathbf{k}} \mu_C W_{\mathbf{k}}^\dagger &=& \mu_C.
\label{eq.Wprop}
\end{eqnarray}
It is immediately clear that this is starkly different from the fermionic case. In particular, the matrix that diagonalizes $H_{\mathbf{k}}^b$ does \textit{not} diagonalize $(H_{\mathbf{k}}^b)^2$. Instead, it diagonalizes $(H_{\mathbf{k}}^b \mu_CH_{\mathbf{k}}^b)$. This can be seen as follows,
\begin{eqnarray} 
\nonumber &\phantom{a}& W_{\mathbf{k}}^\dagger (H_{\mathbf{k}}^b \mu_CH_{\mathbf{k}}^b)W_{\mathbf{k}} = W_{\mathbf{k}}^\dagger H_{\mathbf{k}}^b W_{\mathbf{k}} \mu_C W_{\mathbf{k}}^\dagger H_{\mathbf{k}}^b W_{\mathbf{k}} \\
\nonumber  &=& \mathrm{Diag}\{ E_1,E_2,E_3,E_4\}\times \mu_C \times \mathrm{Diag}\{ E_1,E_2,E_3,E_4\}  \\
 &=& \mathrm{Diag}\{ E_1^2,E_2^2,-E_3^2,-E_4^2\}.
 \label{eq.HsHevals}
\end{eqnarray}
Here, we have replaced $\mu_C$ with $W_{\mathbf{k}} \mu_C W_{\mathbf{k}}^\dagger$ using Eq.~\ref{eq.Wprop}. This can be rephrased as follows: the \textit{bosonic} eigenvalues (those obtained by a pseudo-unitary transformation) of $ (H_{\mathbf{k}}^b \mu_CH_{\mathbf{k}}^b)$ are related to those of $H_{\mathbf{k}}^b$ by the above simple relation. This can be compared with the fermionic case wherein the \textit{fermionic} eigenvalues (those obtained by a unitary transformation) of $H_{\mathbf{k}}^2$ are related to those of $H_{\mathbf{k}}$ as shown in Eq.~\ref{eq.Hsquare}. As in the fermionic case, if we are able to determine the \textit{bosonic} eigenvalues of $(H_{\mathbf{k}}^b \mu_CH_{\mathbf{k}}^b)$ by inspection, we can easily deduce those of $H_\mathbf{k}^b$.

\subsection{Algebra of matrices}

In analogy with the fermionic problem, we note that the Hamiltonian for magnons in the honeycomb XY ferromagnet is spanned by a set of matrices,
\begin{eqnarray}
H_\mathbf{k}^b = g_0 (\mathbf{k}) \mathbb{1}   
+ g_x (\mathbf{k})\mu_x + g_y(\mathbf{k}) \mu_y + g_z(\mathbf{k}) \mu_z,
\label{eq.BosonSpin}
\end{eqnarray}
where $\mu_\alpha = (\sigma_{0}-\sigma_{x}) \otimes \sigma_{\alpha}$ with $\alpha=x,y,z$. 
Here, we find that $g_0(\mathbf{k}) = 3+h/J$, $g_x(\mathbf{k}) = \mathrm{Re} \phantom{a}\epsilon(\mathbf{k})$, $g_y(\mathbf{k}) = \mathrm{Im}\phantom{a}\epsilon(\mathbf{k})$ and $g_z(\mathbf{k}) = 0$. 
Remarkably, this set of matrices forms an analogue of a Clifford algebra. This can be seen from 
\begin{eqnarray}
H_{\mathbf{k}}^b \mu_CH_{\mathbf{k}}^b = \sum_{\alpha,\beta \in \{0,x,y,z\}} g_\alpha g_{\beta} (1-\frac{\delta_{\alpha\beta}}{2}) [\mu_\alpha,\mu_{\beta}]_{\mu_C},
\label{eq.HsigmaH}
\end{eqnarray}
where we have defined an operation between two matrices, $[C,D]_{\mu_C}\equiv ( C \mu_C D + D\mu_C C)$, in analogy with the anti-commutation operation. Here, $\mu_0$ denotes the identity matrix, $\mathbb{1}$. We see that the matrices in Eq.~\ref{eq.BosonSpin} satisfy the following properties,
\begin{eqnarray}
\nonumber [\mathbb{1} ,\mathbb{1} ]_{\mu_C} &=& 2\mu_C, \\
\nonumber [\mu_\alpha,\mu_\alpha]_{\mu_C} &=& 0; \phantom{abc}\alpha=x,y,z \\
\nonumber [\mathbb{1},\mu_\alpha]_{\mu_C} &=& 2(\sigma_z \otimes \sigma_\alpha); \phantom{ac}\alpha=x,y,z \\
\phantom{a} [ \mu_\alpha,\mu_\beta]_{\mu_C} &=& 0; \phantom{ab} \alpha \neq \beta;  \alpha,\beta=x,y,z.
\label{eq.CAconditions}
\end{eqnarray}
Using these properties in Eq.~\ref{eq.HsigmaH}, we find 
\begin{eqnarray}
H_{\mathbf{k}}^b \mu_C H_{\mathbf{k}}^b =  g_0^2 \mu_C + \sum_{a=x,y,z} 2g_0 g_a (\sigma_z \otimes \sigma_a).
\end{eqnarray}
The bosonic eigenvalues of this matrix can be easily found by a simple pseudo-unitary transformation 
using the matrix $\tilde{W} = (\sigma_z \otimes u)$, where $u$ is a $2\times2$ unitary matrix such that $u^\dagger ( 2g_0 g_\alpha \sigma_\alpha) u$ is diagonal. 
This transformation preserves bosonic commutations as it satisfies $\tilde{W} \mu_C \tilde{W}^\dagger = 
\tilde{W}^\dagger \mu_C \tilde{W} = \mu_C$. This is easily seen from rewriting the commutation matrix as $\mu_C = \sigma_z \otimes \sigma_0$.
The matrix $\tilde{W}$ also diagonalizes the matrix given above. As $u$ is a unitary matrix which diagonalizes a $2\times2$ Hermitian matrix spanned by a (fermionic) Clifford algebra, we have 
$u^\dagger ( 2g_0 \sum_a g_a \sigma_a) u =2g_0  \mathrm{Diag}\{+\vert \mathbf{g} \vert,-\vert \mathbf{g} \vert\}$, where $\vert \mathbf{g} \vert = \sqrt{g_x^2 + g_y^2 + g_z^2}$.

This leads to a simple form for the eigenvalues, 
\begin{eqnarray}
\nonumber &\phantom{a}& \tilde{W}^\dagger H_{\mathbf{k}} \mu_C H_{\mathbf{k}} \tilde{W} = g_0^2 \mu_C + 2g_0 (\sigma_z \otimes \mathrm{Diag}\{+\vert  \mathbf{g}\vert ,-\vert \mathbf{g} \vert \}) \\
\nonumber &=& \!\! \left(\!\!\begin{array}{cccc}
g_0^2 +2g_0 \vert \mathbf{g} \vert & & & \\
& g_0^2 -2g_0 \vert \mathbf{g} \vert & & \\
& &  -g_0^2 -2g_0 \vert \mathbf{g} \vert & \\
& & & -g_0^2 +2g_0 \vert \mathbf{g} \vert
\end{array}\!\!\right).
\end{eqnarray}
Comparing this with Eq.~\ref{eq.HsHevals}, we can easily read off the eigenvalues of the original bosonic Hamiltonian $H_\mathbf{k}^b$ as $E_1 = E_3 = \sqrt{g_0^2 + 2g_0 \vert \mathbf{g} \vert}$ and $E_2 = E_4 = \sqrt{g_0^2 - 2g_0 \vert \mathbf{g} \vert}$. 
We have two distinct eigenvalues, forming two bands.

The magnon dispersion for the honeycomb XY model shows a clear cone-like feature at the $K$ point of the Brillouin zone where two bands touch. 
Unlike the case of fermions on the honeycomb lattice, we only have a single $K$ point due to the primed summation in the Hamiltonian. In this vicinity, with $\mathbf{q}$ denoting displacement from the $K$ point, the terms in the Hamiltonian simplify as 
\begin{eqnarray}
g_x (\mathbf{k}) =  \frac{\sqrt{3}}{4} q_{x},\phantom{a} g_y (\mathbf{k}) =  \frac{\sqrt{3}}{4} q_{y}, \phantom{a} g_z(\mathbf{k})=0. \phantom{ac}
\end{eqnarray}
The eigenvalues of the Hamiltonian $H_\mathbf{k}^b$ take a particularly elegant form: $\sqrt{ g_0^2 \pm \frac{g_0\sqrt{3}}{2}  \vert \mathbf{q} \vert}$, where $g_0 = 3+h/J$. This precisely describes a cone-like band touching. To see this, we may Taylor expand the eigenvalues, assuming that the amplitude of $\mathbf{q}$ is small. This leads to $E_\pm(\mathbf{q}) = g_0 \pm \frac{\sqrt{3}}{4 } \vert \mathbf{q}\vert$, with the two bands dispersing linearly to form a cone.

\subsection{Mass term}
As with honeycomb fermions, we have discussed an algebra of three matrices. In the magnonic Hamiltonian for the honeycomb XY model above, the third matrix does not appear as the coefficient $g_z(\mathbf{k})$ vanishes uniformly. This allows us to identify a suitable `mass' term -- a perturbation that will lead to a non-zero value of the $g_z(\mathbf{k})$ at the Dirac point. This takes the form of next-nearest neighbour Y-Y coupling with opposite signs on the A and B sublattices.
\begin{eqnarray}
\nonumber &\phantom{a}& H_{mass} = \Delta \sum_{i\in A} \sum_{\eta} \left[ S_{i,A}^y S_{i+\eta,A}^y - S_{i,B}^y S_{i+\eta,B}^y\right] =\\
\nonumber &{\Delta S}& \sum_{\mathbf{k}}{}^{'} \zeta_\mathbf{k} \left[ a_{\mathbf{k}}^\dagger a_{\mathbf{k}} - a_{\mathbf{k}}^\dagger a_{-\mathbf{k}}^\dagger - a_{-\mathbf{k}} a_{\mathbf{k}} + a_{-\mathbf{k}} a_{-\mathbf{k}}^\dagger - (a\leftrightarrow b)
  \right] \\
  &\rightarrow&   \frac{\Delta}{J} \sum_{\mathbf{k}}{}^{'} \zeta_\mathbf{k}  \mu_z.
\end{eqnarray}
Here, $\eta$ sums over the six next-nearest neighbours of the honeycomb lattice as in the Haldane model. 
We have defined $\zeta_\mathbf{k} = \sum_\eta e^{i\mathbf{k}\cdot \eta}$.
In the Hamiltonian in terms of magnon operators, we obtain the third matrix $\mu_z$, completing the triad of Clifford algebra-like matrices. The resulting dispersion for magnons, in the vicinity of the Dirac cone, is given by
\begin{eqnarray}
E_{mass} \approx  \sqrt{ g_0^2 \pm \frac{g_0\sqrt{3}}{2} \sqrt{q_x^2 + q_y^2 +12\frac{\Delta^2}{J^2}} }.
\end{eqnarray}
Taking the mass term to be small, we may expand the eigenvalues in powers of $\Delta$. At the Dirac point, the eigenvalues come out to be $ g_0 \pm \frac{3\Delta}{2J} + \mathcal{O}(\Delta^2)$.
We have a band gap that is proportional to $\vert \Delta \vert$.

\subsection{Energetic arguments for a mass term}
In fermionic systems with Dirac cones, there is strong tendency to develop a mass term and to  open a gap. For instance, this can arise from a quadratic decoupling of a two-particle interaction term. The propensity towards a mass term can be seen from the following energetic argument. A mass term will open the largest possible gap at the Dirac point. In turn, this will lead to states being pushed furthest down below the Fermi level, leading to maximal energy lowering. This argument underlies the utility of Clifford algebras and mass terms. 

With suitable modifications, a similar argument can be put forth for the above-defined bosonic Clifford algebras. As the excitations are bosonic, there is no Fermi level. However, in systems such as magnets, the eigenvalues of spin-wave excitations contribute to the semi-classical correction to the ground state energy. Constructing the spin wave theory as an expansion in powers of $S$, we have
\begin{eqnarray}
E_{GS} \sim E_{Classical} (S^2)  + \sum_{\mathbf{k}}{}^{'} \sum_{j}\omega_{j,\mathbf{k}} (S) +\ldots.
\label{eq.Eclsp}
\end{eqnarray}  
The first term is the classical ground state energy, proportional to $S^2$. The next term, of order $S$, is given by the sum of spin wave energies over half the Brillouin zone. The index $j$ sums over all spin wave bands.  This form arises from taking an expectation value of the Hamiltonian, e.g., that given by Eq.~\ref{eq.Hamiltonianbosons}. The terms neglected in Eq.~\ref{eq.Eclsp} above, denoted by `$\ldots$', include constants and terms subleading in powers of $S$.

\begin{figure}
\includegraphics[width=3.25in]{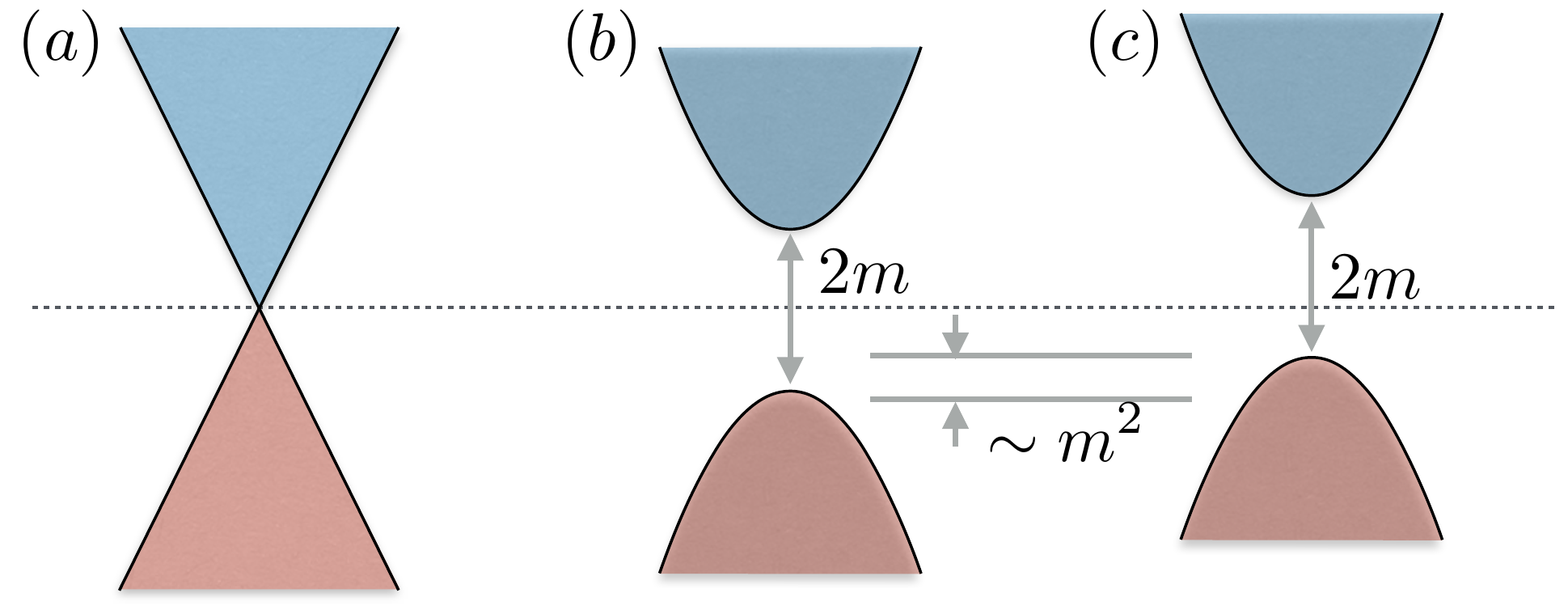}
\caption{(a) Dirac cone in the magnon spectrum of the honeycomb XY model. (b) Gapped dispersion obtained by introducing a mass term in the form of a Y-Y coupling. The gap is taken to be $2m$. (c) Dispersion obtained by introducing a DM interaction so as to open a gap of $2m$. Apart from opening a gap, this shifts both the bands upwards.}
\label{fig.toymodel}
\end{figure}

In a Dirac cone (with only $\mu_x$ and $\mu_y$ in the Hamiltonian, say), we see that states both above and below the Dirac cone contribute to the $\mathcal{O}(S)$ energy. The spin wave eigenvalues, in the vicinity of the Dirac cone, are given by $E_{Dirac}\sim g_0 \pm c \vert \mathbf{q} \vert$. This is shown in Fig.~\ref{fig.toymodel} (left). The $\mathcal{O}(S)$ energy, after adding contributions from both bands, is simply $g_0$ per momentum-space-point. 
Now, if a mass term, proportional to $\mu_z$, is introduced, it changes the eigenvalues to $E_{mass}\sim g_0 \pm c \sqrt{ \vert \mathbf{q} \vert^2 + m^2 }$, where $m$ is proportional to the strength of the mass term at the Dirac point. This changes the energies above and below the Dirac cone in a symmetric fashion. The lower band is pushed down and the upper band is pushed up by the same amount. The sum over energies in the second term of Eq.~\ref{eq.Eclsp} remains at $g_0$ per momentum-space-point. Thus, the mass term does not change the energy to $\mathcal{O}(S)$. 

As opposed to mass terms, we may also introduce other perturbations to open a gap. For example, Ref.~\onlinecite{Owerre2016} shows that a gap can be opened in the honeycomb XY model using a next-nearest neighbour Dzyaloshinskii-Moriya (DM) interaction as shown in Fig.~\ref{fig.lattice}. This term does not fall within our Clifford algebra paradigm; nevertheless it opens a gap. The resulting eigenvalues are
\begin{eqnarray}
E_{DM} \approx J \sqrt{ g_0^2 + 27 D^2 \pm \frac{\sqrt{3}g_0}{2} \sqrt{q_x^2 + q_y^2 + 144 D^2} },
\end{eqnarray}
where $D$ is the magnitude of the DM interaction.
However, this term increases the $\mathcal{O}(S)$ ground state energy. This is because the new DM interaction, apart from opening a gap, shifts both the bands upwards by an amount proportional to $D^2$. This is shown in Fig.~\ref{fig.toymodel}. The sum over all $\omega_{j,\mathbf{k}}$ in Eq.~\ref{eq.Eclsp} then increases, leading to a higher energy cost as compared to the mass term.
 We have checked that other ways (with non-mass terms) to open a band gap generically shift the bands upward. Therefore, we argue that introducing a mass term opens a gap with the least energy $\mathcal{O}(S)$ energy cost. 
 
\section{Topological band structure in one dimension}
\label{sec.triplon}
\begin{figure}
\includegraphics[width=3in]{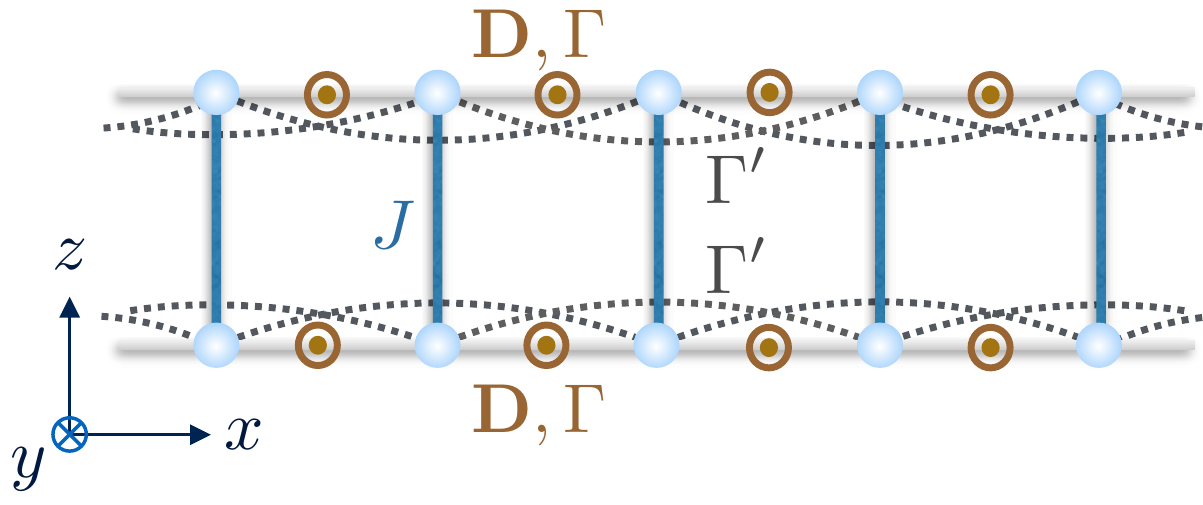}
\caption{Two leg ladder to realize a winding number topological invariant.}
\label{fig.ladder}
\end{figure}

In this section, we show how the new Clifford algebra structure can provide a simple route to topology in one dimension, by presenting an analogy with polyacetylene. We consider a two-leg ladder, shown in Fig.~\ref{fig.ladder}, that is similar to the model discussed in Ref.~\onlinecite{Joshi2017}. We have a two-site unit cell labelled by the index $\ell=1,2$.
The rung bonds are assumed to be antiferromagnetic, providing the dominant energy scale in the problem. The inter-rung bonds are assumed to only have DM interactions and symmetric off-diagonal exchange. The DM interactions are assumed to point in the (negative) Y direction. The symmetric off-diagonal exchange is also taken to be in the Y direction. In addition, we consider next nearest neighbour symmetric off-diagonal coupling, $\Gamma'$; we show below that this serves as a tuning knob to induce a topological phase transition. The Hamiltonian is given by
\begin{eqnarray}
\nonumber H_{l} &=& J\sum_{i}\mathbf{S}_{i,1}\cdot\mathbf{S}_{i,2}
+\Gamma \sum_{i,\ell=1,2} \left[ S_{i,\ell}^x S_{i+1,\ell}^z + S_{i,\ell}^z S_{i+1,\ell}^x \right]\\
\nonumber &+&D \sum_{i,\ell=1,2} \left[ S_{i,\ell}^x S_{i+1,\ell}^z - S_{i,\ell}^z S_{i+1,\ell}^x \right]\\
&+&\Gamma' \sum_{i,\ell=1,2} \left[ S_{i,\ell}^x S_{i+2,\ell}^z + S_{i,\ell}^z S_{i+2,\ell}^x \right].
\label{eq.tripH}
\end{eqnarray} 
When $J$ is the largest scale, such Hamiltonians give rise to gapped dimerized ground states characterized by singlet formation on each rung. The excitations correspond to breaking singlets to create triplets leading to `triplon' quasiparticles. The bond operator prescription of Sachdev and Bhatt\cite{Sachdev1990} can be used to describe these excitations. We introduce a bosonic representation on each dimer with $\vert s\rangle = s^\dagger \vert 0 \rangle = \frac{1}{\sqrt{2}} \{\vert \!\! \uparrow \downarrow\rangle - \vert  \!\!\downarrow \uparrow\rangle\}$, $\vert t_x\rangle = t_x^\dagger \vert 0 \rangle = \frac{i}{\sqrt{2}} \{\vert\!\! \uparrow \uparrow\rangle - \vert  \!\!\downarrow \downarrow\rangle\}$,  
$\vert t_y\rangle = t_y^\dagger \vert 0 \rangle = \frac{1}{\sqrt{2}} \{\vert\!\! \uparrow \uparrow\rangle + \vert\!\!  \downarrow \downarrow\rangle\}$ and
$\vert t_z\rangle = t_z^\dagger \vert 0 \rangle = \frac{-i}{\sqrt{2}} \{\vert\!\! \uparrow \downarrow\rangle + \vert  \!\!\downarrow \uparrow\rangle\}$.
Here, $\vert 0 \rangle$ is an unphysical vacuum. In order to remain in the physical subspace, we must have an occupancy of one boson per site. In the spirit of mean field theory, we will satisfy this constraint on average by introducing a chemical potential, $\mu$. To describe the dimerized state, we take the singlet boson to be Bose condensed with $s_i^\dagger \approx s_i \approx \bar{s}$. 
The intra-rung Hamiltonian is given by
\begin{equation}
H_{J} = \sum_{i}  \left( 
\frac{-3J\bar{s}^2}{4} -\mu \bar{s}^2  +\left\{\frac{J}{4}-\mu\right\}\sum_{\alpha = x,y,z} t_{i,\alpha}^\dagger t_{i,\alpha}
\right).
\label{eq.HJ}
\end{equation}
The parameters $\bar{s}$, $\mu$ can be obtained self-consistently. However, in the following analysis, their precise values are not important. 
The inter-rung couplings can be expressed using $
S_{i,\ell=1,2}^{\alpha=x,y,z} = \frac{i\bar{s}}{2} \{ t_{i,\alpha}^\dagger - t_{i,\alpha}\} + \mathcal{O}(\bar{s}^0)$. Expanding the resulting Hamiltonian in powers of $\bar{s}$, the leading terms are quadratic in the triplon operators. The $t_y$ bosons decouple from the other two species. Focussing on $t_x$ and $t_z$ alone, we obtain
\begin{eqnarray}
H_{trip} = \frac{1}{2} \sum_{k}\Psi_{k}^\dagger \mathcal{M}_{k} \Psi_{k} + c,
\end{eqnarray}
where $\Psi_{k} = \left(   
\begin{array}{cccc}
t_{{k},x} & t_{{k},z} & t_{-{k},x}^\dagger & t_{-{k},z}^\dagger
\end{array}
\right)^T$. The constant $c$ depends on parameters $\bar{s}$ and $\mu$.
As before, if ${k}$ is included in the sum, $-{k}$ could be excluded so that we only consider half the Brillouin zone. Nevertheless, we keep the full Brillouin zone and include a factor of $1/2$ in the Hamiltonian to account for double counting. 

Surprisingly, the Hamiltonian matrix $\mathcal{M}_{k}$ can be expressed in terms of the bosonic Clifford algebra matrices introduced above. We find
\begin{eqnarray}
\mathcal{M}_\mathbf{k} = h_0 \mathbb{1} + h_x (k)\mu_x + h_y (k)\mu_y, 
\label{eq.tripmumats}
\end{eqnarray}
where $\mathbb{1} = \sigma_0 \otimes \sigma_0$, $\mu_x = (\sigma_0-\sigma_x) \otimes \sigma_x $ and $\mu_y =  (\sigma_0-\sigma_x) \otimes \sigma_y$, as defined below Eq.~\ref{eq.BosonSpin} above. 
The coefficients are given by $h_0 = (J/4-\mu)$, $h_x(k) = \Gamma \bar{s}^2 \cos k +\Gamma' \bar{s}^2 \cos 2k$ and $h_y (k)= -D \bar{s}^2 \sin k$.
Following our earlier analysis, the eigenvalues of the Hamiltonian can be immediately read off as $
\sqrt{h_0^2 \pm 2 h_0\sqrt{h_x^2 (k)+ h_y^2 (k)}}$.
These eigenvalues represent two bands which are separated by a band gap if $h_x(k)$ and $h_y(k)$ are non-zero for all $k$. For example, if $\Gamma'$ is turned off, we immediately see that the band gap survives as long as $\Gamma$ and $D$ are non-zero.

\begin{figure}
\includegraphics[width=3.in]{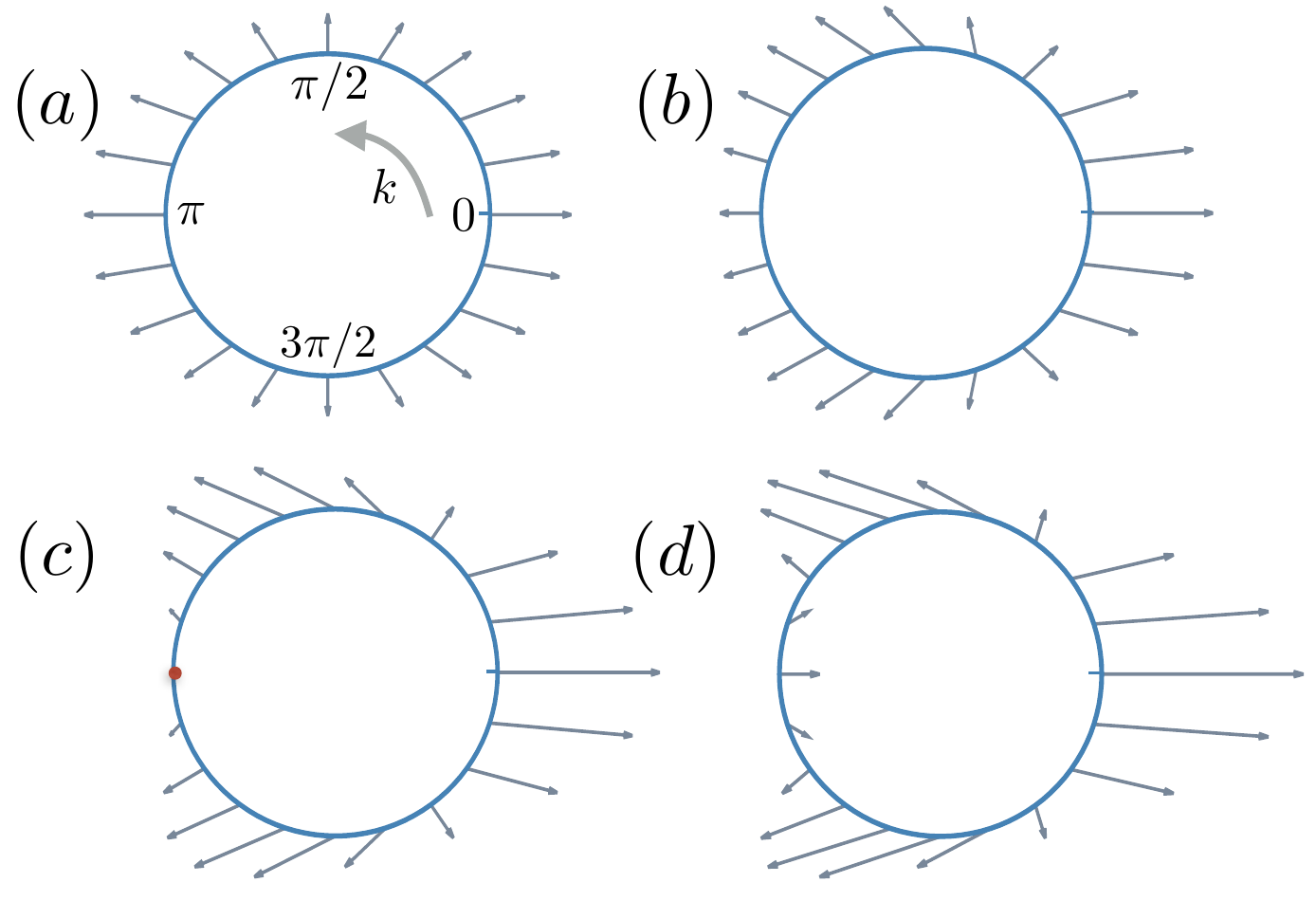}
\caption{Winding of $\vec{h}$, the vector encoding the triplon Hamiltonian for $\Gamma'/\Gamma = 0, 0.5,1,1.5$ (a-d). The blue circles represent the Brillouin zone, with $k$ running from $0$ to $2\pi$. The grey arrows represent $\vec{h}$ for different $k$ values as we move along the Brillouin zone. When $\Gamma'=\Gamma$, $\vec{h}$ vanishes when $k=\pi$, indicated by a red dot. 
We have fixed $\Gamma=1$ and $D=0.5$.
}
\label{fig.winding}
\end{figure}

More importantly, Eq.~\ref{eq.tripmumats} brings out the topological character of the Hamiltonian, providing a bosonic analogue of the celebrated Su-Schrieffer-Heeger model for polyacetylene\cite{Heeger1988}. We see that the Hamiltonian serves as a map from ${S}^1$ (the Brillouin zone) to $\mathbb{R}^2$ (the linear space spanned by coefficients of $\mu_x$ and $\mu_y$). This mapping can occur in topologically distinct sectors which are indexed by the winding number. To see this, we take $\Gamma'$ to be a tuning parameter as we examine the variation of the vector $\vec{h}=\{\Gamma \bar{s}^2 \cos k + \Gamma' \bar{s}^2 \cos 2k, -D \bar{s}^2 \sin k,0\}$ as we tune $k$ from $-\pi$ to $\pi$. 
As shown in Fig.~\ref{fig.winding}(a), when $\Gamma'=0$,  $\vec{h}$ is readily seen to wind once counter-clockwise as we go around the Brillouin zone. This winding character remains robust as $\Gamma'/\Gamma$ is increased as shown in Fig.~\ref{fig.winding}(b). When $\Gamma'=\Gamma$, we see that $h_x$ and $h_y$ vanish at $k=\pi$, as shown in Fig.~\ref{fig.winding}(c). At this point, the winding number is not well-defined as the band gap closes and a topological phase transition occurs. Upon increasing $\Gamma'/\Gamma$ further as in Fig.~\ref{fig.winding}(d), the winding number becomes zero.

\begin{figure}
\includegraphics[width=3.in]{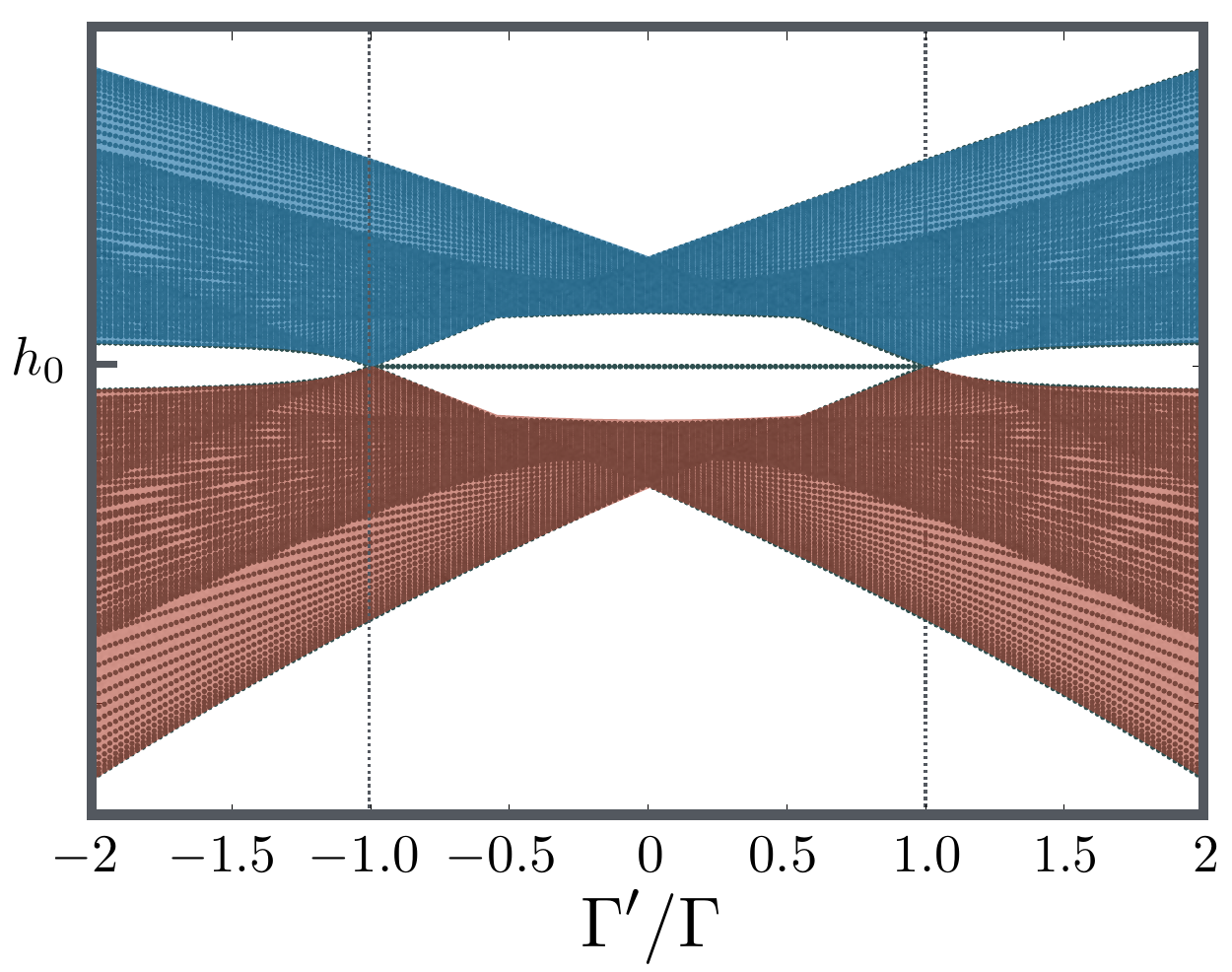}
\caption{Eigenenergies on an open ladder vs. $\Gamma'/\Gamma$. Mid gap states with energy $h_0$ appear for $-\Gamma < \Gamma'  < \Gamma$. These states are all localized at the edges. Topological transitions at $\Gamma' = \pm \Gamma$ are shown as dotted vertical lines. 
}
\label{fig.energies}
\end{figure}

The winding number illustrated in Fig.~\ref{fig.winding} is a topological invariant which cannot be changed without closing the band gap. A non-zero winding indicates a topologically non-trivial phase. In the presence of an edge, the topological character leads to protected edge states as shown in Fig.~\ref{fig.energies}. In the topological regime, we find that an open configuration gives rise to two mid-gap states with energy $h_0$. They are both localized at the two edges.

\section{Discussion}
We have presented analogues of the Dirac equation and of Clifford algebras that are applicable in bosonic systems. 
While bosonic topological phases have been gaining interest, there is no rigorous way to classify their band structures. Indeed, the celebrated ten-fold 
classification\cite{Altland1997,Schnyder2008,Ludwig2016} does not apply to bosonic systems due to non-compactness of the eigenvector space. In this light, we hope that our discovery of a Clifford-like algebra may help to understand possible topological phases of bosons. 

As a systematic topological classification is not available for bosons, earlier works have extrapolated fermionic invariants (such as winding numbers\cite{Joshi2017} and Chern numbers\cite{Shindou2013}) to bosonic band structures. In some cases where pairing terms in the Hamiltonian turn out to be numerically small, it has been argued that the pseudo-unitary nature of the eigenvectors can be neglected\cite{Romhanyi2015}. The eigenvectors then become equivalent to that of a fermionic problem. In such a case, the Hamiltonian can be approximated as a topological mapping to a closed orientable surface that may or may not enclose the origin\cite{Romhanyi2015,Romhanyi2018}. Here, we have presented a Clifford algebra paradigm that provides a direct geometric understanding of topology in bosonic systems. As we have shown with the ladder problem, quantities such as the winding number can be directly deduced. 

We have presented the conditions required for a bosonic Clifford algebra in Eqs.~\ref{eq.CAconditions}. We have explicitly constructed three matrices $\{ \mu_{\alpha = x,y,z}=(\sigma_0 - \sigma_x)\otimes \sigma_\alpha\}$ that satisfy these requirements. These matrices naturally emerge in the honeycomb XY model and in the ladder problem that we have discussed.
Among $4\times4$ matrices, there are three other triads that also satisfy the conditions in  Eqs.~\ref{eq.CAconditions}. We have $\{ \lambda_{\alpha}=(\sigma_0 + \sigma_x)\otimes \sigma_\alpha\}$,  $\{ \nu_{\alpha}=(\sigma_0 + \sigma_y)\otimes \sigma_\alpha\}$, $\{ \xi_{\alpha}=(\sigma_0 - \sigma_y)\otimes \sigma_\alpha\}$. These triads may be realized in other bosonic systems with Dirac cones. As with fermions\cite{Herbut2012}, when considering matrices with dimension greater than four, we may find larger sets of matrices that satisfy this algebra.

Our definition of the bosonic Clifford algebra places strong constraints on the Hamiltonian and on its topology. Consider a $4\times4$ Hamiltonian in two dimensions,
\begin{eqnarray}
H_\mathbf{k} = g_0 + g_x (\mathbf{k}) \mu_x+g_y (\mathbf{k}) \mu_y + g_z (\mathbf{k}) \mu_z, 
\end{eqnarray}
where $\mu_{x,y,z}$ are as defined above (see below Eq.~\ref{eq.BosonSpin}). As the Hamiltonian has a primed summation over $\mathbf{k}$, we need not consider $\vec{g}(-\mathbf{k})$ if we have included $\mathbf{k}$ in our summation. Nevertheless, in order to examine topology, we consider the vector $\vec{g}(\mathbf{k})$ defined over the entire Brillouin zone.
From the form of the Hamiltonian, we see that $\mu_x$ and $\mu_z$ are symmetric under $\mathbf{k}\rightarrow -\mathbf{k}$, while $\mu_y$ is antisymmetric. For example, $\mu_{z}$ encodes terms such as $\{a_{\mathbf{k}}^\dagger a_{\mathbf{k}} + a_{-\mathbf{k}} a_{-\mathbf{k}}^\dagger \}$, while $\mu_y$ encodes $\{ i a_{\mathbf{k}}^\dagger b_{\mathbf{k}} - i b_{-\mathbf{k}} a_{-\mathbf{k}}^\dagger \}$.

This constrains $g_x$ and $g_z$ to be even functions of $\mathbf{k}$ while $g_y$ must be an odd function. If not, the contribution from these terms would vanish when summed over the full Brillouin zone. In this light, we now examine the topology inherent in the Hamiltonian which constitutes a mapping from the Brillouin zone (a two dimensional torus) to $\mathbb{R}^3$, the space of three-component vectors. We may expect this mapping to have topological sectors characterized by skyrmion number, given by
\begin{equation}
N_S = \int dk_x \int dk_y  \left\{ \frac{\partial \vec{g}}{\partial k_x} \times \frac{\partial \vec{g}}{\partial k_y} \right\}\cdot \vec{g},
\end{equation}
where the integral is over the full Brillouin zone. This quantity necessarily \textit{vanishes} due to the properties of $g_{x,y,z}$ under inversion. As the y-component of $\vec{g}$ alone flips sign under inversion, the contributions from patches centred around $\mathbf{k}$ and $-\mathbf{k}$ cancel each other.
Even if we were to restrict the integral to half the Brillouin zone, the resulting skyrmion number would depend on the precise choice of $\mathbf{k}$-points (the half-Brillouin zone). As a result, it cannot constitute a topological invariant. This shows that, unlike in fermionic systems, a two-dimensional Hamiltonian spanned by the bosonic Clifford algebra cannot have non-trivial skyrmion sectors. An exciting future direction is to see if this algebra allows for higher topological notions such as $\mathbb{Z}_2$ invariants.

\acknowledgments We are thankful to R. Shankar (Yale) for encouraging discussions.

\bibliographystyle{apsrev4-1} 
\bibliography{bosonic_clifford}

\begin{thebibliography}{40}%
\makeatletter
\providecommand \@ifxundefined [1]{%
 \@ifx{#1\undefined}
}%
\providecommand \@ifnum [1]{%
 \ifnum #1\expandafter \@firstoftwo
 \else \expandafter \@secondoftwo
 \fi
}%
\providecommand \@ifx [1]{%
 \ifx #1\expandafter \@firstoftwo
 \else \expandafter \@secondoftwo
 \fi
}%
\providecommand \natexlab [1]{#1}%
\providecommand \enquote  [1]{``#1''}%
\providecommand \bibnamefont  [1]{#1}%
\providecommand \bibfnamefont [1]{#1}%
\providecommand \citenamefont [1]{#1}%
\providecommand \href@noop [0]{\@secondoftwo}%
\providecommand \href [0]{\begingroup \@sanitize@url \@href}%
\providecommand \@href[1]{\@@startlink{#1}\@@href}%
\providecommand \@@href[1]{\endgroup#1\@@endlink}%
\providecommand \@sanitize@url [0]{\catcode `\\12\catcode `\$12\catcode
  `\&12\catcode `\#12\catcode `\^12\catcode `\_12\catcode `\%12\relax}%
\providecommand \@@startlink[1]{}%
\providecommand \@@endlink[0]{}%
\providecommand \url  [0]{\begingroup\@sanitize@url \@url }%
\providecommand \@url [1]{\endgroup\@href {#1}{\urlprefix }}%
\providecommand \urlprefix  [0]{URL }%
\providecommand \Eprint [0]{\href }%
\providecommand \doibase [0]{http://dx.doi.org/}%
\providecommand \selectlanguage [0]{\@gobble}%
\providecommand \bibinfo  [0]{\@secondoftwo}%
\providecommand \bibfield  [0]{\@secondoftwo}%
\providecommand \translation [1]{[#1]}%
\providecommand \BibitemOpen [0]{}%
\providecommand \bibitemStop [0]{}%
\providecommand \bibitemNoStop [0]{.\EOS\space}%
\providecommand \EOS [0]{\spacefactor3000\relax}%
\providecommand \BibitemShut  [1]{\csname bibitem#1\endcsname}%
\let\auto@bib@innerbib\@empty
\bibitem [{\citenamefont {Onose}\ \emph {et~al.}(2010)\citenamefont {Onose},
  \citenamefont {Ideue}, \citenamefont {Katsura}, \citenamefont {Shiomi},
  \citenamefont {Nagaosa},\ and\ \citenamefont {Tokura}}]{Onose2010}%
  \BibitemOpen
  \bibfield  {author} {\bibinfo {author} {\bibfnamefont {Y.}~\bibnamefont
  {Onose}}, \bibinfo {author} {\bibfnamefont {T.}~\bibnamefont {Ideue}},
  \bibinfo {author} {\bibfnamefont {H.}~\bibnamefont {Katsura}}, \bibinfo
  {author} {\bibfnamefont {Y.}~\bibnamefont {Shiomi}}, \bibinfo {author}
  {\bibfnamefont {N.}~\bibnamefont {Nagaosa}}, \ and\ \bibinfo {author}
  {\bibfnamefont {Y.}~\bibnamefont {Tokura}},\ }\href {\doibase
  10.1126/science.1188260} {\bibfield  {journal} {\bibinfo  {journal}
  {Science}\ }\textbf {\bibinfo {volume} {329}},\ \bibinfo {pages} {297}
  (\bibinfo {year} {2010})},\ \Eprint
  {http://arxiv.org/abs/http://www.sciencemag.org/content/329/5989/297.full.pdf}
  {http://www.sciencemag.org/content/329/5989/297.full.pdf} \BibitemShut
  {NoStop}%
\bibitem [{\citenamefont {Matsumoto}\ and\ \citenamefont
  {Murakami}(2011)}]{Matsumoto2011}%
  \BibitemOpen
  \bibfield  {author} {\bibinfo {author} {\bibfnamefont {R.}~\bibnamefont
  {Matsumoto}}\ and\ \bibinfo {author} {\bibfnamefont {S.}~\bibnamefont
  {Murakami}},\ }\href {\doibase 10.1103/PhysRevLett.106.197202} {\bibfield
  {journal} {\bibinfo  {journal} {Phys. Rev. Lett.}\ }\textbf {\bibinfo
  {volume} {106}},\ \bibinfo {pages} {197202} (\bibinfo {year}
  {2011})}\BibitemShut {NoStop}%
\bibitem [{\citenamefont {van Hoogdalem}\ \emph {et~al.}(2013)\citenamefont
  {van Hoogdalem}, \citenamefont {Tserkovnyak},\ and\ \citenamefont
  {Loss}}]{Hoogdalem2013}%
  \BibitemOpen
  \bibfield  {author} {\bibinfo {author} {\bibfnamefont {K.~A.}\ \bibnamefont
  {van Hoogdalem}}, \bibinfo {author} {\bibfnamefont {Y.}~\bibnamefont
  {Tserkovnyak}}, \ and\ \bibinfo {author} {\bibfnamefont {D.}~\bibnamefont
  {Loss}},\ }\href {\doibase 10.1103/PhysRevB.87.024402} {\bibfield  {journal}
  {\bibinfo  {journal} {Phys. Rev. B}\ }\textbf {\bibinfo {volume} {87}},\
  \bibinfo {pages} {024402} (\bibinfo {year} {2013})}\BibitemShut {NoStop}%
\bibitem [{\citenamefont {Zhang}\ \emph {et~al.}(2013)\citenamefont {Zhang},
  \citenamefont {Ren}, \citenamefont {Wang},\ and\ \citenamefont
  {Li}}]{Zhang2013}%
  \BibitemOpen
  \bibfield  {author} {\bibinfo {author} {\bibfnamefont {L.}~\bibnamefont
  {Zhang}}, \bibinfo {author} {\bibfnamefont {J.}~\bibnamefont {Ren}}, \bibinfo
  {author} {\bibfnamefont {J.-S.}\ \bibnamefont {Wang}}, \ and\ \bibinfo
  {author} {\bibfnamefont {B.}~\bibnamefont {Li}},\ }\href {\doibase
  10.1103/PhysRevB.87.144101} {\bibfield  {journal} {\bibinfo  {journal} {Phys.
  Rev. B}\ }\textbf {\bibinfo {volume} {87}},\ \bibinfo {pages} {144101}
  (\bibinfo {year} {2013})}\BibitemShut {NoStop}%
\bibitem [{\citenamefont {Romh{\'a}nyi}\ \emph {et~al.}(2015)\citenamefont
  {Romh{\'a}nyi}, \citenamefont {Penc},\ and\ \citenamefont
  {Ganesh}}]{Romhanyi2015}%
  \BibitemOpen
  \bibfield  {author} {\bibinfo {author} {\bibfnamefont {J.}~\bibnamefont
  {Romh{\'a}nyi}}, \bibinfo {author} {\bibfnamefont {K.}~\bibnamefont {Penc}},
  \ and\ \bibinfo {author} {\bibfnamefont {R.}~\bibnamefont {Ganesh}},\ }\href
  {http://dx.doi.org/10.1038/ncomms7805} {\bibfield  {journal} {\bibinfo
  {journal} {Nature Communications}\ }\textbf {\bibinfo {volume} {6}},\
  \bibinfo {pages} {6805 EP } (\bibinfo {year} {2015})}\BibitemShut {NoStop}%
\bibitem [{\citenamefont {Haldane}(1988)}]{Haldane1988}%
  \BibitemOpen
  \bibfield  {author} {\bibinfo {author} {\bibfnamefont {F.~D.~M.}\
  \bibnamefont {Haldane}},\ }\href {\doibase 10.1103/PhysRevLett.61.2015}
  {\bibfield  {journal} {\bibinfo  {journal} {Phys. Rev. Lett.}\ }\textbf
  {\bibinfo {volume} {61}},\ \bibinfo {pages} {2015} (\bibinfo {year}
  {1988})}\BibitemShut {NoStop}%
\bibitem [{\citenamefont {Kane}\ and\ \citenamefont
  {Mele}(2005)}]{KaneMele2005}%
  \BibitemOpen
  \bibfield  {author} {\bibinfo {author} {\bibfnamefont {C.~L.}\ \bibnamefont
  {Kane}}\ and\ \bibinfo {author} {\bibfnamefont {E.~J.}\ \bibnamefont
  {Mele}},\ }\href {\doibase 10.1103/PhysRevLett.95.226801} {\bibfield
  {journal} {\bibinfo  {journal} {Phys. Rev. Lett.}\ }\textbf {\bibinfo
  {volume} {95}},\ \bibinfo {pages} {226801} (\bibinfo {year}
  {2005})}\BibitemShut {NoStop}%
\bibitem [{\citenamefont {Janssen}\ and\ \citenamefont
  {Herbut}(2015)}]{Janssen2015}%
  \BibitemOpen
  \bibfield  {author} {\bibinfo {author} {\bibfnamefont {L.}~\bibnamefont
  {Janssen}}\ and\ \bibinfo {author} {\bibfnamefont {I.~F.}\ \bibnamefont
  {Herbut}},\ }\href {\doibase 10.1103/PhysRevB.92.045117} {\bibfield
  {journal} {\bibinfo  {journal} {Phys. Rev. B}\ }\textbf {\bibinfo {volume}
  {92}},\ \bibinfo {pages} {045117} (\bibinfo {year} {2015})}\BibitemShut
  {NoStop}%
\bibitem [{\citenamefont {Moon}(2016)}]{Moon2016}%
  \BibitemOpen
  \bibfield  {author} {\bibinfo {author} {\bibfnamefont {E.-G.}\ \bibnamefont
  {Moon}},\ }\href {http://dx.doi.org/10.1038/srep31051} {\bibfield  {journal}
  {\bibinfo  {journal} {Scientific Reports}\ }\textbf {\bibinfo {volume} {6}},\
  \bibinfo {pages} {31051 EP } (\bibinfo {year} {2016})}\BibitemShut {NoStop}%
\bibitem [{\citenamefont {{Huang}}\ \emph {et~al.}(2012)\citenamefont
  {{Huang}}, \citenamefont {{Liu}},\ and\ \citenamefont {{Chan}}}]{Huang2012}%
  \BibitemOpen
  \bibfield  {author} {\bibinfo {author} {\bibfnamefont {X.}~\bibnamefont
  {{Huang}}}, \bibinfo {author} {\bibfnamefont {F.}~\bibnamefont {{Liu}}}, \
  and\ \bibinfo {author} {\bibfnamefont {C.~T.}\ \bibnamefont {{Chan}}},\
  }\href@noop {} {\bibfield  {journal} {\bibinfo  {journal} {ArXiv e-prints}\ }
  (\bibinfo {year} {2012})},\ \Eprint {http://arxiv.org/abs/1205.0886}
  {arXiv:1205.0886 [cond-mat.mtrl-sci]} \BibitemShut {NoStop}%
\bibitem [{\citenamefont {Yu}\ \emph {et~al.}(2015)\citenamefont {Yu},
  \citenamefont {Wang}, \citenamefont {Zheng}, \citenamefont {He},
  \citenamefont {Liu}, \citenamefont {Lu},\ and\ \citenamefont
  {Chen}}]{Yu2015}%
  \BibitemOpen
  \bibfield  {author} {\bibinfo {author} {\bibfnamefont {S.-Y.}\ \bibnamefont
  {Yu}}, \bibinfo {author} {\bibfnamefont {Q.}~\bibnamefont {Wang}}, \bibinfo
  {author} {\bibfnamefont {L.-Y.}\ \bibnamefont {Zheng}}, \bibinfo {author}
  {\bibfnamefont {C.}~\bibnamefont {He}}, \bibinfo {author} {\bibfnamefont
  {X.-P.}\ \bibnamefont {Liu}}, \bibinfo {author} {\bibfnamefont {M.-H.}\
  \bibnamefont {Lu}}, \ and\ \bibinfo {author} {\bibfnamefont {Y.-F.}\
  \bibnamefont {Chen}},\ }\href {\doibase 10.1063/1.4918651} {\bibfield
  {journal} {\bibinfo  {journal} {Applied Physics Letters}\ }\textbf {\bibinfo
  {volume} {106}},\ \bibinfo {pages} {151906} (\bibinfo {year} {2015})},\
  \Eprint {http://arxiv.org/abs/https://doi.org/10.1063/1.4918651}
  {https://doi.org/10.1063/1.4918651} \BibitemShut {NoStop}%
\bibitem [{\citenamefont {Yu}\ \emph {et~al.}(2016)\citenamefont {Yu},
  \citenamefont {Sun}, \citenamefont {Ni}, \citenamefont {Wang}, \citenamefont
  {Yan}, \citenamefont {He}, \citenamefont {Liu}, \citenamefont {Feng},
  \citenamefont {Lu},\ and\ \citenamefont {Chen}}]{Yu2016}%
  \BibitemOpen
  \bibfield  {author} {\bibinfo {author} {\bibfnamefont {S.-Y.}\ \bibnamefont
  {Yu}}, \bibinfo {author} {\bibfnamefont {X.-C.}\ \bibnamefont {Sun}},
  \bibinfo {author} {\bibfnamefont {X.}~\bibnamefont {Ni}}, \bibinfo {author}
  {\bibfnamefont {Q.}~\bibnamefont {Wang}}, \bibinfo {author} {\bibfnamefont
  {X.-J.}\ \bibnamefont {Yan}}, \bibinfo {author} {\bibfnamefont
  {C.}~\bibnamefont {He}}, \bibinfo {author} {\bibfnamefont {X.-P.}\
  \bibnamefont {Liu}}, \bibinfo {author} {\bibfnamefont {L.}~\bibnamefont
  {Feng}}, \bibinfo {author} {\bibfnamefont {M.-H.}\ \bibnamefont {Lu}}, \ and\
  \bibinfo {author} {\bibfnamefont {Y.-F.}\ \bibnamefont {Chen}},\ }\href
  {http://dx.doi.org/10.1038/nmat4743} {\bibfield  {journal} {\bibinfo
  {journal} {Nature Materials}\ }\textbf {\bibinfo {volume} {15}},\ \bibinfo
  {pages} {1243 EP } (\bibinfo {year} {2016})}\BibitemShut {NoStop}%
\bibitem [{\citenamefont {Gao}\ \emph {et~al.}(2016)\citenamefont {Gao},
  \citenamefont {Zhang}, \citenamefont {Wu}, \citenamefont {Yao},\ and\
  \citenamefont {Li}}]{Gao2016}%
  \BibitemOpen
  \bibfield  {author} {\bibinfo {author} {\bibfnamefont {H.-F.}\ \bibnamefont
  {Gao}}, \bibinfo {author} {\bibfnamefont {X.}~\bibnamefont {Zhang}}, \bibinfo
  {author} {\bibfnamefont {F.-G.}\ \bibnamefont {Wu}}, \bibinfo {author}
  {\bibfnamefont {Y.-W.}\ \bibnamefont {Yao}}, \ and\ \bibinfo {author}
  {\bibfnamefont {J.}~\bibnamefont {Li}},\ }\href {\doibase
  https://doi.org/10.1016/j.ssc.2016.03.002} {\bibfield  {journal} {\bibinfo
  {journal} {Solid State Communications}\ }\textbf {\bibinfo {volume}
  {234-235}},\ \bibinfo {pages} {35 } (\bibinfo {year} {2016})}\BibitemShut
  {NoStop}%
\bibitem [{\citenamefont {Sakoda}(2012)}]{Sakoda2012}%
  \BibitemOpen
  \bibfield  {author} {\bibinfo {author} {\bibfnamefont {K.}~\bibnamefont
  {Sakoda}},\ }\href {\doibase 10.1364/OE.20.003898} {\bibfield  {journal}
  {\bibinfo  {journal} {Opt. Express}\ }\textbf {\bibinfo {volume} {20}},\
  \bibinfo {pages} {3898} (\bibinfo {year} {2012})}\BibitemShut {NoStop}%
\bibitem [{\citenamefont {Chan}\ \emph {et~al.}(2012)\citenamefont {Chan},
  \citenamefont {Hang},\ and\ \citenamefont {Huang}}]{Chan2012}%
  \BibitemOpen
  \bibfield  {author} {\bibinfo {author} {\bibfnamefont {C.~T.}\ \bibnamefont
  {Chan}}, \bibinfo {author} {\bibfnamefont {Z.~H.}\ \bibnamefont {Hang}}, \
  and\ \bibinfo {author} {\bibfnamefont {X.}~\bibnamefont {Huang}},\ }\href
  {\doibase doi:10.1155/2012/313984} {\bibfield  {journal} {\bibinfo  {journal}
  {Advances in OptoElectronics}\ }\textbf {\bibinfo {volume} {2012}},\ \bibinfo
  {pages} {313984} (\bibinfo {year} {2012})}\BibitemShut {NoStop}%
\bibitem [{\citenamefont {He}\ and\ \citenamefont {Chan}(2015)}]{He2015}%
  \BibitemOpen
  \bibfield  {author} {\bibinfo {author} {\bibfnamefont {W.-Y.}\ \bibnamefont
  {He}}\ and\ \bibinfo {author} {\bibfnamefont {C.~T.}\ \bibnamefont {Chan}},\
  }\href {http://dx.doi.org/10.1038/srep08186} {\bibfield  {journal} {\bibinfo
  {journal} {Scientific Reports}\ }\textbf {\bibinfo {volume} {5}},\ \bibinfo
  {pages} {8186 EP } (\bibinfo {year} {2015})}\BibitemShut {NoStop}%
\bibitem [{\citenamefont {Yi}\ and\ \citenamefont {Karzig}(2016)}]{Yi2016}%
  \BibitemOpen
  \bibfield  {author} {\bibinfo {author} {\bibfnamefont {K.}~\bibnamefont
  {Yi}}\ and\ \bibinfo {author} {\bibfnamefont {T.}~\bibnamefont {Karzig}},\
  }\href {\doibase 10.1103/PhysRevB.93.104303} {\bibfield  {journal} {\bibinfo
  {journal} {Phys. Rev. B}\ }\textbf {\bibinfo {volume} {93}},\ \bibinfo
  {pages} {104303} (\bibinfo {year} {2016})}\BibitemShut {NoStop}%
\bibitem [{\citenamefont {Owerre}(2016)}]{Owerre2016}%
  \BibitemOpen
  \bibfield  {author} {\bibinfo {author} {\bibfnamefont {S.~A.}\ \bibnamefont
  {Owerre}},\ }\href {http://stacks.iop.org/0953-8984/28/i=38/a=386001}
  {\bibfield  {journal} {\bibinfo  {journal} {Journal of Physics: Condensed
  Matter}\ }\textbf {\bibinfo {volume} {28}},\ \bibinfo {pages} {386001}
  (\bibinfo {year} {2016})}\BibitemShut {NoStop}%
\bibitem [{\citenamefont {Fransson}\ \emph {et~al.}(2016)\citenamefont
  {Fransson}, \citenamefont {Black-Schaffer},\ and\ \citenamefont
  {Balatsky}}]{Fransson2016}%
  \BibitemOpen
  \bibfield  {author} {\bibinfo {author} {\bibfnamefont {J.}~\bibnamefont
  {Fransson}}, \bibinfo {author} {\bibfnamefont {A.~M.}\ \bibnamefont
  {Black-Schaffer}}, \ and\ \bibinfo {author} {\bibfnamefont {A.~V.}\
  \bibnamefont {Balatsky}},\ }\href {\doibase 10.1103/PhysRevB.94.075401}
  {\bibfield  {journal} {\bibinfo  {journal} {Phys. Rev. B}\ }\textbf {\bibinfo
  {volume} {94}},\ \bibinfo {pages} {075401} (\bibinfo {year}
  {2016})}\BibitemShut {NoStop}%
\bibitem [{\citenamefont {Kim}\ \emph {et~al.}(2016)\citenamefont {Kim},
  \citenamefont {Ochoa}, \citenamefont {Zarzuela},\ and\ \citenamefont
  {Tserkovnyak}}]{Kim2016}%
  \BibitemOpen
  \bibfield  {author} {\bibinfo {author} {\bibfnamefont {S.~K.}\ \bibnamefont
  {Kim}}, \bibinfo {author} {\bibfnamefont {H.}~\bibnamefont {Ochoa}}, \bibinfo
  {author} {\bibfnamefont {R.}~\bibnamefont {Zarzuela}}, \ and\ \bibinfo
  {author} {\bibfnamefont {Y.}~\bibnamefont {Tserkovnyak}},\ }\href {\doibase
  10.1103/PhysRevLett.117.227201} {\bibfield  {journal} {\bibinfo  {journal}
  {Phys. Rev. Lett.}\ }\textbf {\bibinfo {volume} {117}},\ \bibinfo {pages}
  {227201} (\bibinfo {year} {2016})}\BibitemShut {NoStop}%
\bibitem [{\citenamefont {Owerre}(2017)}]{Owerre_SciRep_2017}%
  \BibitemOpen
  \bibfield  {author} {\bibinfo {author} {\bibfnamefont {S.~A.}\ \bibnamefont
  {Owerre}},\ }\href {\doibase 10.1038/s41598-017-07276-8} {\bibfield
  {journal} {\bibinfo  {journal} {Scientific Reports}\ }\textbf {\bibinfo
  {volume} {7}},\ \bibinfo {pages} {6931} (\bibinfo {year} {2017})}\BibitemShut
  {NoStop}%
\bibitem [{\citenamefont {McClarty}\ \emph {et~al.}(2017)\citenamefont
  {McClarty}, \citenamefont {Kr{\"u}ger}, \citenamefont {Guidi}, \citenamefont
  {Parker}, \citenamefont {Refson}, \citenamefont {Parker}, \citenamefont
  {Prabhakaran},\ and\ \citenamefont {Coldea}}]{McClarty2017}%
  \BibitemOpen
  \bibfield  {author} {\bibinfo {author} {\bibfnamefont {P.~A.}\ \bibnamefont
  {McClarty}}, \bibinfo {author} {\bibfnamefont {F.}~\bibnamefont
  {Kr{\"u}ger}}, \bibinfo {author} {\bibfnamefont {T.}~\bibnamefont {Guidi}},
  \bibinfo {author} {\bibfnamefont {S.~F.}\ \bibnamefont {Parker}}, \bibinfo
  {author} {\bibfnamefont {K.}~\bibnamefont {Refson}}, \bibinfo {author}
  {\bibfnamefont {A.~W.}\ \bibnamefont {Parker}}, \bibinfo {author}
  {\bibfnamefont {D.}~\bibnamefont {Prabhakaran}}, \ and\ \bibinfo {author}
  {\bibfnamefont {R.}~\bibnamefont {Coldea}},\ }\href
  {http://dx.doi.org/10.1038/nphys4117} {\bibfield  {journal} {\bibinfo
  {journal} {Nature Physics}\ }\textbf {\bibinfo {volume} {13}},\ \bibinfo
  {pages} {736 EP } (\bibinfo {year} {2017})}\BibitemShut {NoStop}%
\bibitem [{\citenamefont {Joshi}\ and\ \citenamefont
  {Schnyder}(2017)}]{Joshi2017}%
  \BibitemOpen
  \bibfield  {author} {\bibinfo {author} {\bibfnamefont {D.~G.}\ \bibnamefont
  {Joshi}}\ and\ \bibinfo {author} {\bibfnamefont {A.~P.}\ \bibnamefont
  {Schnyder}},\ }\href {\doibase 10.1103/PhysRevB.96.220405} {\bibfield
  {journal} {\bibinfo  {journal} {Phys. Rev. B}\ }\textbf {\bibinfo {volume}
  {96}},\ \bibinfo {pages} {220405} (\bibinfo {year} {2017})}\BibitemShut
  {NoStop}%
\bibitem [{\citenamefont {Li}\ \emph {et~al.}(2016)\citenamefont {Li},
  \citenamefont {Li}, \citenamefont {Kim}, \citenamefont {Balents},
  \citenamefont {Yu},\ and\ \citenamefont {Chen}}]{Li2016}%
  \BibitemOpen
  \bibfield  {author} {\bibinfo {author} {\bibfnamefont {F.-Y.}\ \bibnamefont
  {Li}}, \bibinfo {author} {\bibfnamefont {Y.-D.}\ \bibnamefont {Li}}, \bibinfo
  {author} {\bibfnamefont {Y.~B.}\ \bibnamefont {Kim}}, \bibinfo {author}
  {\bibfnamefont {L.}~\bibnamefont {Balents}}, \bibinfo {author} {\bibfnamefont
  {Y.}~\bibnamefont {Yu}}, \ and\ \bibinfo {author} {\bibfnamefont
  {G.}~\bibnamefont {Chen}},\ }\href {http://dx.doi.org/10.1038/ncomms12691}
  {\bibfield  {journal} {\bibinfo  {journal} {Nature Communications}\ }\textbf
  {\bibinfo {volume} {7}},\ \bibinfo {pages} {12691 EP } (\bibinfo {year}
  {2016})}\BibitemShut {NoStop}%
\bibitem [{\citenamefont {Mook}\ \emph {et~al.}(2016)\citenamefont {Mook},
  \citenamefont {Henk},\ and\ \citenamefont {Mertig}}]{Mook2016}%
  \BibitemOpen
  \bibfield  {author} {\bibinfo {author} {\bibfnamefont {A.}~\bibnamefont
  {Mook}}, \bibinfo {author} {\bibfnamefont {J.}~\bibnamefont {Henk}}, \ and\
  \bibinfo {author} {\bibfnamefont {I.}~\bibnamefont {Mertig}},\ }\href
  {\doibase 10.1103/PhysRevLett.117.157204} {\bibfield  {journal} {\bibinfo
  {journal} {Phys. Rev. Lett.}\ }\textbf {\bibinfo {volume} {117}},\ \bibinfo
  {pages} {157204} (\bibinfo {year} {2016})}\BibitemShut {NoStop}%
\bibitem [{\citenamefont {Li}\ \emph {et~al.}(2017{\natexlab{a}})\citenamefont
  {Li}, \citenamefont {Li}, \citenamefont {Hu}, \citenamefont {Li},\ and\
  \citenamefont {Fang}}]{Kangkang2017}%
  \BibitemOpen
  \bibfield  {author} {\bibinfo {author} {\bibfnamefont {K.}~\bibnamefont
  {Li}}, \bibinfo {author} {\bibfnamefont {C.}~\bibnamefont {Li}}, \bibinfo
  {author} {\bibfnamefont {J.}~\bibnamefont {Hu}}, \bibinfo {author}
  {\bibfnamefont {Y.}~\bibnamefont {Li}}, \ and\ \bibinfo {author}
  {\bibfnamefont {C.}~\bibnamefont {Fang}},\ }\href {\doibase
  10.1103/PhysRevLett.119.247202} {\bibfield  {journal} {\bibinfo  {journal}
  {Phys. Rev. Lett.}\ }\textbf {\bibinfo {volume} {119}},\ \bibinfo {pages}
  {247202} (\bibinfo {year} {2017}{\natexlab{a}})}\BibitemShut {NoStop}%
\bibitem [{\citenamefont {Li}\ \emph {et~al.}(2017{\natexlab{b}})\citenamefont
  {Li}, \citenamefont {Li}, \citenamefont {Yu}, \citenamefont {Paramekanti},\
  and\ \citenamefont {Chen}}]{Li2017}%
  \BibitemOpen
  \bibfield  {author} {\bibinfo {author} {\bibfnamefont {F.-Y.}\ \bibnamefont
  {Li}}, \bibinfo {author} {\bibfnamefont {Y.-D.}\ \bibnamefont {Li}}, \bibinfo
  {author} {\bibfnamefont {Y.}~\bibnamefont {Yu}}, \bibinfo {author}
  {\bibfnamefont {A.}~\bibnamefont {Paramekanti}}, \ and\ \bibinfo {author}
  {\bibfnamefont {G.}~\bibnamefont {Chen}},\ }\href {\doibase
  10.1103/PhysRevB.95.085132} {\bibfield  {journal} {\bibinfo  {journal} {Phys.
  Rev. B}\ }\textbf {\bibinfo {volume} {95}},\ \bibinfo {pages} {085132}
  (\bibinfo {year} {2017}{\natexlab{b}})}\BibitemShut {NoStop}%
\bibitem [{\citenamefont {Su}\ and\ \citenamefont {Wang}(2017)}]{Ying2017}%
  \BibitemOpen
  \bibfield  {author} {\bibinfo {author} {\bibfnamefont {Y.}~\bibnamefont
  {Su}}\ and\ \bibinfo {author} {\bibfnamefont {X.~R.}\ \bibnamefont {Wang}},\
  }\href {\doibase 10.1103/PhysRevB.96.104437} {\bibfield  {journal} {\bibinfo
  {journal} {Phys. Rev. B}\ }\textbf {\bibinfo {volume} {96}},\ \bibinfo
  {pages} {104437} (\bibinfo {year} {2017})}\BibitemShut {NoStop}%
\bibitem [{\citenamefont {{Jian}}\ and\ \citenamefont
  {{Nie}}(2017)}]{Jian2017}%
  \BibitemOpen
  \bibfield  {author} {\bibinfo {author} {\bibfnamefont {S.-K.}\ \bibnamefont
  {{Jian}}}\ and\ \bibinfo {author} {\bibfnamefont {W.}~\bibnamefont {{Nie}}},\
  }\href@noop {} {\bibfield  {journal} {\bibinfo  {journal} {ArXiv e-prints}\ }
  (\bibinfo {year} {2017})},\ \Eprint {http://arxiv.org/abs/1708.02948}
  {arXiv:1708.02948 [cond-mat.str-el]} \BibitemShut {NoStop}%
\bibitem [{\citenamefont {Semenoff}(1984)}]{Semenoff1984}%
  \BibitemOpen
  \bibfield  {author} {\bibinfo {author} {\bibfnamefont {G.~W.}\ \bibnamefont
  {Semenoff}},\ }\href {\doibase 10.1103/PhysRevLett.53.2449} {\bibfield
  {journal} {\bibinfo  {journal} {Phys. Rev. Lett.}\ }\textbf {\bibinfo
  {volume} {53}},\ \bibinfo {pages} {2449} (\bibinfo {year}
  {1984})}\BibitemShut {NoStop}%
\bibitem [{\citenamefont {Hunt}\ \emph {et~al.}(2013)\citenamefont {Hunt},
  \citenamefont {Sanchez-Yamagishi}, \citenamefont {Young}, \citenamefont
  {Yankowitz}, \citenamefont {LeRoy}, \citenamefont {Watanabe}, \citenamefont
  {Taniguchi}, \citenamefont {Moon}, \citenamefont {Koshino}, \citenamefont
  {Jarillo-Herrero},\ and\ \citenamefont {Ashoori}}]{Hunt2013}%
  \BibitemOpen
  \bibfield  {author} {\bibinfo {author} {\bibfnamefont {B.}~\bibnamefont
  {Hunt}}, \bibinfo {author} {\bibfnamefont {J.~D.}\ \bibnamefont
  {Sanchez-Yamagishi}}, \bibinfo {author} {\bibfnamefont {A.~F.}\ \bibnamefont
  {Young}}, \bibinfo {author} {\bibfnamefont {M.}~\bibnamefont {Yankowitz}},
  \bibinfo {author} {\bibfnamefont {B.~J.}\ \bibnamefont {LeRoy}}, \bibinfo
  {author} {\bibfnamefont {K.}~\bibnamefont {Watanabe}}, \bibinfo {author}
  {\bibfnamefont {T.}~\bibnamefont {Taniguchi}}, \bibinfo {author}
  {\bibfnamefont {P.}~\bibnamefont {Moon}}, \bibinfo {author} {\bibfnamefont
  {M.}~\bibnamefont {Koshino}}, \bibinfo {author} {\bibfnamefont
  {P.}~\bibnamefont {Jarillo-Herrero}}, \ and\ \bibinfo {author} {\bibfnamefont
  {R.~C.}\ \bibnamefont {Ashoori}},\ }\href {\doibase 10.1126/science.1237240}
  {\bibfield  {journal} {\bibinfo  {journal} {Science}\ }\textbf {\bibinfo
  {volume} {340}},\ \bibinfo {pages} {1427} (\bibinfo {year} {2013})},\ \Eprint
  {http://arxiv.org/abs/http://science.sciencemag.org/content/340/6139/1427.full.pdf}
  {http://science.sciencemag.org/content/340/6139/1427.full.pdf} \BibitemShut
  {NoStop}%
\bibitem [{\citenamefont {Blaizot}\ and\ \citenamefont
  {Ripka}(1986)}]{Blaizot1986}%
  \BibitemOpen
  \bibfield  {author} {\bibinfo {author} {\bibfnamefont {J.}~\bibnamefont
  {Blaizot}}\ and\ \bibinfo {author} {\bibfnamefont {G.}~\bibnamefont
  {Ripka}},\ }\href {https://books.google.co.in/books?id=s\_xlQgAACAAJ} {\emph
  {\bibinfo {title} {Quantum Theory of Finite Systems}}}\ (\bibinfo
  {publisher} {Cambridge, MA},\ \bibinfo {year} {1986})\BibitemShut {NoStop}%
\bibitem [{\citenamefont {Sachdev}\ and\ \citenamefont
  {Bhatt}(1990)}]{Sachdev1990}%
  \BibitemOpen
  \bibfield  {author} {\bibinfo {author} {\bibfnamefont {S.}~\bibnamefont
  {Sachdev}}\ and\ \bibinfo {author} {\bibfnamefont {R.~N.}\ \bibnamefont
  {Bhatt}},\ }\href {\doibase 10.1103/PhysRevB.41.9323} {\bibfield  {journal}
  {\bibinfo  {journal} {Phys. Rev. B}\ }\textbf {\bibinfo {volume} {41}},\
  \bibinfo {pages} {9323} (\bibinfo {year} {1990})}\BibitemShut {NoStop}%
\bibitem [{\citenamefont {Heeger}\ \emph {et~al.}(1988)\citenamefont {Heeger},
  \citenamefont {Kivelson}, \citenamefont {Schrieffer},\ and\ \citenamefont
  {Su}}]{Heeger1988}%
  \BibitemOpen
  \bibfield  {author} {\bibinfo {author} {\bibfnamefont {A.~J.}\ \bibnamefont
  {Heeger}}, \bibinfo {author} {\bibfnamefont {S.}~\bibnamefont {Kivelson}},
  \bibinfo {author} {\bibfnamefont {J.~R.}\ \bibnamefont {Schrieffer}}, \ and\
  \bibinfo {author} {\bibfnamefont {W.~P.}\ \bibnamefont {Su}},\ }\href
  {\doibase 10.1103/RevModPhys.60.781} {\bibfield  {journal} {\bibinfo
  {journal} {Rev. Mod. Phys.}\ }\textbf {\bibinfo {volume} {60}},\ \bibinfo
  {pages} {781} (\bibinfo {year} {1988})}\BibitemShut {NoStop}%
\bibitem [{\citenamefont {Altland}\ and\ \citenamefont
  {Zirnbauer}(1997)}]{Altland1997}%
  \BibitemOpen
  \bibfield  {author} {\bibinfo {author} {\bibfnamefont {A.}~\bibnamefont
  {Altland}}\ and\ \bibinfo {author} {\bibfnamefont {M.~R.}\ \bibnamefont
  {Zirnbauer}},\ }\href {\doibase 10.1103/PhysRevB.55.1142} {\bibfield
  {journal} {\bibinfo  {journal} {Phys. Rev. B}\ }\textbf {\bibinfo {volume}
  {55}},\ \bibinfo {pages} {1142} (\bibinfo {year} {1997})}\BibitemShut
  {NoStop}%
\bibitem [{\citenamefont {Schnyder}\ \emph {et~al.}(2008)\citenamefont
  {Schnyder}, \citenamefont {Ryu}, \citenamefont {Furusaki},\ and\
  \citenamefont {Ludwig}}]{Schnyder2008}%
  \BibitemOpen
  \bibfield  {author} {\bibinfo {author} {\bibfnamefont {A.~P.}\ \bibnamefont
  {Schnyder}}, \bibinfo {author} {\bibfnamefont {S.}~\bibnamefont {Ryu}},
  \bibinfo {author} {\bibfnamefont {A.}~\bibnamefont {Furusaki}}, \ and\
  \bibinfo {author} {\bibfnamefont {A.~W.~W.}\ \bibnamefont {Ludwig}},\ }\href
  {\doibase 10.1103/PhysRevB.78.195125} {\bibfield  {journal} {\bibinfo
  {journal} {Phys. Rev. B}\ }\textbf {\bibinfo {volume} {78}},\ \bibinfo
  {pages} {195125} (\bibinfo {year} {2008})}\BibitemShut {NoStop}%
\bibitem [{\citenamefont {Ludwig}(2016)}]{Ludwig2016}%
  \BibitemOpen
  \bibfield  {author} {\bibinfo {author} {\bibfnamefont {A.~W.~W.}\
  \bibnamefont {Ludwig}},\ }\href
  {http://stacks.iop.org/1402-4896/2016/i=T168/a=014001} {\bibfield  {journal}
  {\bibinfo  {journal} {Physica Scripta}\ }\textbf {\bibinfo {volume} {2016}},\
  \bibinfo {pages} {014001} (\bibinfo {year} {2016})}\BibitemShut {NoStop}%
\bibitem [{\citenamefont {Herbut}(2012)}]{Herbut2012}%
  \BibitemOpen
  \bibfield  {author} {\bibinfo {author} {\bibfnamefont {I.~F.}\ \bibnamefont
  {Herbut}},\ }\href {\doibase 10.1103/PhysRevB.85.085304} {\bibfield
  {journal} {\bibinfo  {journal} {Phys. Rev. B}\ }\textbf {\bibinfo {volume}
  {85}},\ \bibinfo {pages} {085304} (\bibinfo {year} {2012})}\BibitemShut
  {NoStop}%
\bibitem [{\citenamefont {Shindou}\ \emph {et~al.}(2013)\citenamefont
  {Shindou}, \citenamefont {Matsumoto}, \citenamefont {Murakami},\ and\
  \citenamefont {Ohe}}]{Shindou2013}%
  \BibitemOpen
  \bibfield  {author} {\bibinfo {author} {\bibfnamefont {R.}~\bibnamefont
  {Shindou}}, \bibinfo {author} {\bibfnamefont {R.}~\bibnamefont {Matsumoto}},
  \bibinfo {author} {\bibfnamefont {S.}~\bibnamefont {Murakami}}, \ and\
  \bibinfo {author} {\bibfnamefont {J.-i.}\ \bibnamefont {Ohe}},\ }\href
  {\doibase 10.1103/PhysRevB.87.174427} {\bibfield  {journal} {\bibinfo
  {journal} {Phys. Rev. B}\ }\textbf {\bibinfo {volume} {87}},\ \bibinfo
  {pages} {174427} (\bibinfo {year} {2013})}\BibitemShut {NoStop}%
\bibitem [{\citenamefont {{Romh{\'a}nyi}}(2018)}]{Romhanyi2018}%
  \BibitemOpen
  \bibfield  {author} {\bibinfo {author} {\bibfnamefont {J.}~\bibnamefont
  {{Romh{\'a}nyi}}},\ }\href@noop {} {\bibfield  {journal} {\bibinfo  {journal}
  {ArXiv e-prints}\ } (\bibinfo {year} {2018})},\ \Eprint
  {http://arxiv.org/abs/1801.07950} {arXiv:1801.07950 [cond-mat.str-el]}
  \BibitemShut {NoStop}%
\end{thebibliography}%

\end{document}